\documentclass[lettersize,journal]{IEEEtran}
\usepackage{amsmath,amsfonts}
\usepackage{algorithmic}
\usepackage{algorithm}
\usepackage{array}
\usepackage[caption=false,font=footnotesize,labelfont=rm,textfont=rm]{subfig}
\usepackage{textcomp}
\usepackage{stfloats}
\usepackage{url}
\usepackage{verbatim}
\usepackage{graphicx}
\usepackage{cite}
\usepackage{color}
\usepackage{bm}
\usepackage{threeparttable}
\usepackage{makecell}
\usepackage{multirow}

\begin{document}
\title{SCR-Auth: Secure Call Receiver Authentication on Smartphones Using Outer Ear Echoes}

\author{Xiping Sun, Jing Chen, Kun He, Zhixiang He, Ruiying Du, Yebo Feng, Qingchuan Zhao, Cong Wu
	
\thanks{
Xiping Sun, Kun He and Zhixiang He are with Key Laboratory of Aerospace Information Security and Trusted Computing, Ministry of Education, School of Cyber Science and Engineering, Wuhan University, Wuhan, 430072, China (e-mail: \{xiping, hekun, zhixianghe\}@whu.edu.cn).
		
Jing Chen is with Key Laboratory of Aerospace Information Security and Trusted Computing, Ministry of Education, School of Cyber Science and Engineering, Wuhan University, Wuhan, 430072, China, and also with Rizhao Institute of Information Technology, Wuhan University, Rizhao, 276800, China (e-mail: chenjing@whu.edu.cn).

Ruiying Du is with Key Laboratory of Aerospace Information Security and Trusted Computing, Ministry of Education, School of Cyber Science and Engineering, Wuhan University, Wuhan, 430072, China, and also with Collaborative Innovation Center of Geospatial Technology, Wuhan, 430079, China (e-mail: duraying@whu.edu.cn).

Cong Wu and Yebo Feng are with School of Computer Science and Engineering, Nanyang Technological University, Singapore (e-mail: \{cong.wu, yebo.feng\}@ntu.edu.sg).
		
Qingchuan Zhao is with Department of Computer Science, City University of Hong Kong (e-mail: cs.qczhao@cityu.edu.hk).
		
}
}

\maketitle
\begin{abstract}
	Receiving calls is one of the most universal functions of smartphones, involving sensitive information and critical operations.
	Unfortunately, to prioritize convenience, the current call receiving process bypasses smartphone authentication mechanisms (e.g., passwords, fingerprint recognition, and face recognition), leaving a significant security gap.
	To address this issue, we propose SCR-Auth, a secure call receiver authentication scheme for smartphones that leverages outer ear echoes.
	It sends inaudible acoustic signals through the earpiece speaker to actively sense the call receiver's outer ear structure and records the resulting echoes using the top microphone.
	These echoes are then analyzed to extract unique outer ear biometric information for authentication.
	It operates implicitly, without requiring extra hardware or imposing additional burden.
	Comprehensive experiments conducted under diverse conditions demonstrate SCR-Auth’s effectiveness and security, showing an average balanced accuracy of 96.95\% and resilience against potential attacks.
\end{abstract}


\section{Introduction}
\label{sec:introduction}
Phone calls are one of the most widely used and trusted communication forms on smartphones~\cite{pew}, often involving sensitive information and critical operations, such as accessing health records~\cite{privacy} or authorizing financial transactions~\cite{bank}. 
While smartphones have adopted various authentication mechanisms to prevent unauthorized access, including passwords~\cite{moh2024understanding}, fingerprint recognition~\cite{terhorst2021midecon}, and face recognition~\cite{meden2021privacy}, the call receiving process remains an exception. 
Prioritizing convenience, incoming calls bypass these authentications, allowing anyone with physical access to the smartphone to answer even if it is locked, which fails to meet essential security standards. 
Therefore, it is crucial to develop an effective call receiver authentication mechanism that ensures only the legitimate smartphone owner can answer incoming calls while maintaining convenience.

\begin{figure}[!t]
	\centering
	\subfloat[An incoming call]{\includegraphics[width = 0.28\linewidth]{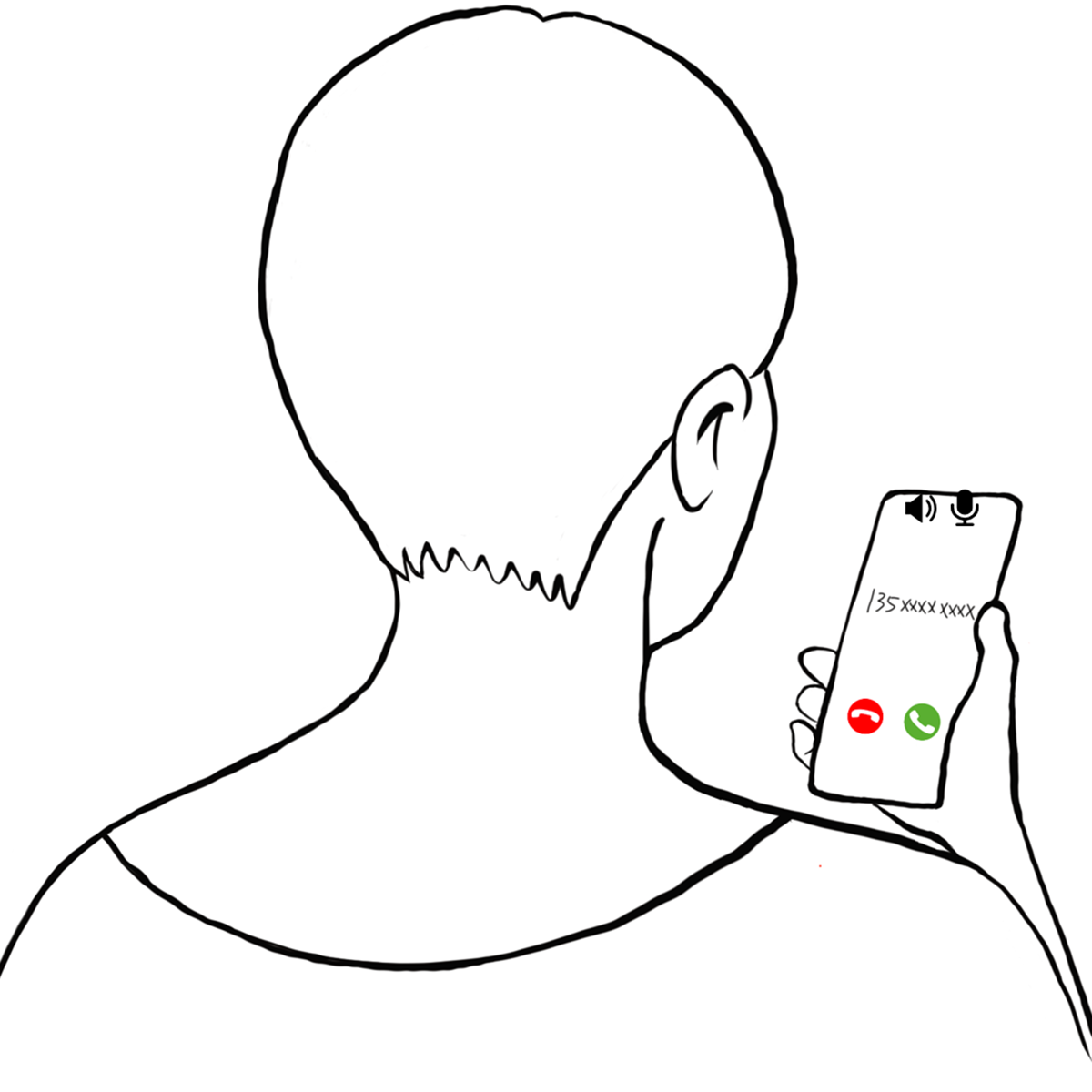}}
	\hspace{1em}
	\subfloat[Call receiver authentication]{\includegraphics[width = 0.5\linewidth]{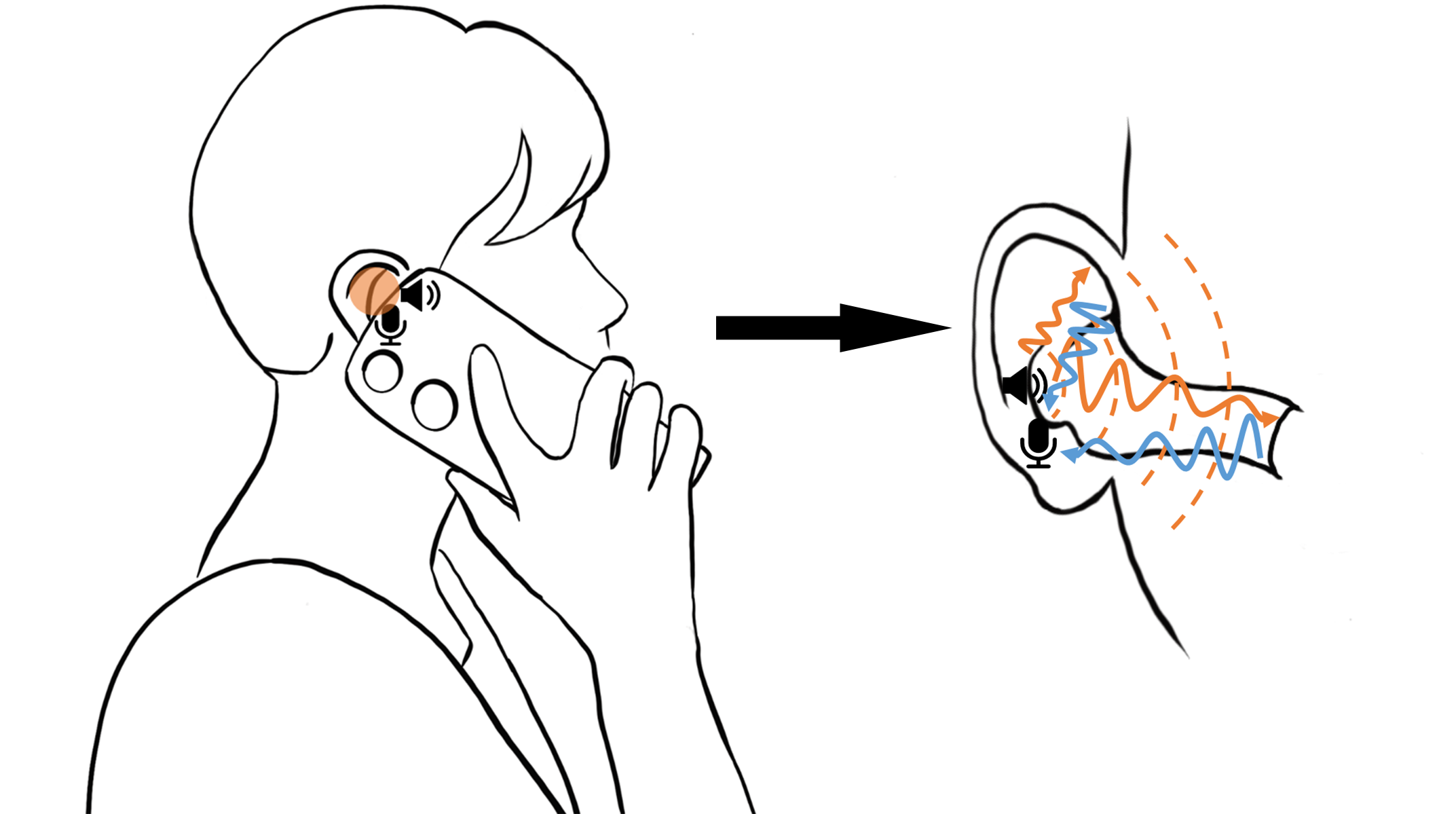}}
	\caption{Illustration of SCR-Auth. When a call comes in, the earpiece speaker and top microphone serve as an active sonar, authenticating the call receiver's identity by analyzing echoes from the outer ear.}
	\label{fig:show}
\end{figure}

There have been several active works that can be applied to call receiver authentication~\cite{Conti2011,lee2017secure,eremin2019touch,buriro2019answerauth,buriro2015itsme,fan2023emgauth,fahmi2012implicit,damer2012ear,Holz2015,cabra2022earprint,itani2022multimodal,mizuho2024earauthcam,Gao2019,wu2023earae}.
Behavioral biometrics-based methods analyze users’ hand motions, such as picking up the smartphone~\cite{Conti2011,lee2017secure,eremin2019touch} or performing sliding gestures~\cite{buriro2019answerauth,buriro2015itsme}. 
However, these methods often suffer from low accuracy due to behavioral variability~\cite{wu2024s}.
Other approaches focus on ear physiological characteristics, capturing ear images with a camera~\cite{fahmi2012implicit,damer2012ear} or pressing the ear against a capacitive touchscreen~\cite{Holz2015,cabra2022earprint}.
However, they require additional user actions and root privileges, and face challenges in low-light conditions.
Recent studies have explored using earphones for user authentication~\cite{mizuho2024earauthcam,Gao2019,wu2023earae}. 
However, these approaches necessitate hardware modifications to existing earphones, such as integrating extra sensors (e.g., cameras or inward-facing microphones), which increases costs and leads to incompatibility with commercial earphones.
Moreover, they require users to constantly carry earphones, significantly reducing convenience and practicality.

In this paper, we propose SCR-Auth, a \underline{s}ecure \underline{c}all \underline{r}eceiver (SCR) authentication method for smartphones based on outer ear echoes.
During the natural call receiving process, SCR-Auth emits acoustic sensing signals through the smartphone’s earpiece speaker, as illustrated in Fig.~\ref{fig:show}.
These signals interact with the user's outer ear, undergoing absorption and reflection before reaching the top microphone.
The resulting echoes carry distinct outer ear biometric information (e.g., auricle shape, ear canal geometry, and tissue properties), which is unique to each individual and can be analyzed for authentication.
SCR-Auth achieves seamless and implicit authentication without requiring extra hardware or imposing additional burden, ensuring a smooth call receiving experience.

Realizing SCR-Auth in practice faces several challenges.
Firstly, due to the multipath effect, the signals captured by the smartphone’s built-in microphone include not only outer ear echoes, but also direct path signals and environmental reflections. 
These signal components overlap in both frequency and phase, making it difficult to effectively filter the interference caused by the direct path signals and environmental reflections.
Secondly, outer ear echoes are sensitive to the relative position between the ear and smartphone due to altered signal propagation properties. 
This sensitivity leads to unstable echo patterns, making reliable feature extraction a challenge.

To address the first challenge, we propose a two-step denoising method. 
The process begins with a bandpass filter to remove ambient noises, followed by the Magnitude-Phase Spectrogram Subtraction (MPSS) method to suppress interference.
Specifically, for each signal segment derived through synchronization and segmentation, we compute both magnitude and phase spectrograms. 
A reference segment is then chosen, which primarily contains direct path signals and environmental reflections, free from outer ear echoes.  
Based on the selected reference segment, we construct differential spectrograms in both the magnitude and phase domains, effectively mitigating unwanted interference.
To counteract the position variability between the ear and smartphone, we design a learning-based feature extractor.
We first train a Convolutional Neural Network (CNN) model using multi-user data collected under diverse natural smartphone positions at call reception.
Through supervised learning, the CNN model is guided to focus on identity-related features while disregarding secondary factors, such as changes in the relative position between the ear and smartphone.
Based on the idea of transfer learning, we then transfer the pre-trained model as a generalized feature extractor to obtain reliable features.
Finally, SCR-Auth adopts a user-specific one-class classification model to verify the legitimacy of the call receiver.

\begin{table*}[!t]
	\centering
	\caption{Comparison of representative call receiver authentication methods on smartphones}
	\label{tab:related}
	\renewcommand\arraystretch{1.01}
	\setlength{\tabcolsep}{1.5mm}{
	\begin{threeparttable}
		\begin{tabular}{c|llcccccc}
			\hline
			Device & System & Distinctiveness & \makecell[c]{No extra \\ hardware\tnote{1}} & \makecell[c]{Little usage \\ constraint\tnote{2}} & \makecell[c]{No root \\ privileges\tnote{3}} & \makecell[c]{Resilient across \\ diverse conditions\tnote{4}} & Accuracy & Error rate \\
			\hline
			\multirow{5}{*}{Smartphones} 
			& Conti \emph{et al.}~\cite{Conti2011}           & Hand movements          & \checkmark & $\times$   & \checkmark & $\times$ & N/A     & $\sim$ 7\% \\
			& Fahmi \emph{et al.}~\cite{fahmi2012implicit}   & Entire ear image            & \checkmark & $\times$   & \checkmark & $\times$ & 92.5\% & N/A \\ 
			& Bodyprint~\cite{Holz2015}                      & Entire ear capacitive image & \checkmark & $\times$   & $\times$   & $\times$ & 99.52\% & 7.8\%        \\
			& Itani \emph{et al.}~\cite{itani2022multimodal} & Image \& Pinna responses    & \checkmark & $\times$   & \checkmark & $\times$ & N/A & 1.6\% \\
			& SCR-Auth (ours)                                & Inaudible outer ear echoes  & \checkmark & \checkmark & \checkmark & \checkmark & 96.95\% & 1.53\%     \\
			\hline
			\multirow{2}{*}{With earphones}
			& EarAuthCam~\cite{mizuho2024earauthcam}         & Upper part of ear image     & $\times$   & \checkmark & \checkmark & \checkmark & 84.1\% & 8.36\% \\
			& EarEcho~\cite{Gao2019}                         & Audible ear canal echoes    & $\times$   & \checkmark & \checkmark & \checkmark & 94.52\% & N/A        \\
			\hline
		\end{tabular}
		\begin{tablenotes}
			\footnotesize
			\item[1] : No extra hardware implies that only commodity smartphones are used, without the need for additional devices or sensors.
			\item[2] : Little usage constraint indicates that there are no requirements on movement patterns, additional gestures or usage environments.
			\item[3] : No root privileges means that there is no need to root the smartphone or modify the kernel source.
			\item[4] : Resilient across diverse conditions means that the method is robust in various situations, such as different environments and user postures.
		\end{tablenotes}
	\end{threeparttable}
}
\end{table*}

In summary, the contributions of this paper are as follows:
\begin{itemize}
	\item
	      We propose SCR-Auth, a novel call receiver authentication scheme for smartphones that leverages outer ear echoes, enabling secure and implicit authentication without the need for extra hardware or user burden.

	\item
	      To eliminate ambient noise, as well as interference from the direct path signal and environmental reflections, we propose a specially designed two-step denoising method, encompassing bandpass filtering and spectrogram differencing.
	      To further enhance system robustness against smartphone position changes, we introduce a pre-trained neural network model that leverages transfer learning to extract reliable features.

	\item
	      We conduct comprehensive experiments under various conditions to evaluate the effectiveness of SCR-Auth, e.g., ambient noises, different postures, different periods, and devices.
	      The results show that SCR-Auth can achieve a balanced accuracy of 96.95\% and a equal error rate of 1.53\%.
	   	  We demonstrate the security of SCR-Auth by evaluating its resistance to common attacks.    	 

\end{itemize}

\section{Related Work}
\label{sec:related Work}
In this section, we review related works on call receiver authentication for smartphones. 
Additionally, we explore recent advancements in the field of acoustic sensing.

\subsection{Call Receiver Authentication}
Authenticating the identity of the call receiver is essential for ensuring both security and privacy on smartphones. 
Call receiver authentication methods can be broadly categorized into two types: behavioral biometrics-based and physiological biometrics-based methods. 
Table~\ref{tab:related} summarizes several representative approaches to call receiver authentication on smartphones.

Behavioral biometrics-based methods authenticate the call receiver by analyzing their behavior during phone call interactions~\cite{Conti2011,lee2017secure,eremin2019touch,buriro2019answerauth,buriro2015itsme,fan2023emgauth}.
These approaches commonly use motion sensors to capture movement patterns, such as how a user picks up the smartphone and positions it to their ear, to verify their identity. 
However, these methods often require users to follow specific movement patterns and suffer from low accuracy due to the inherent variability and uncontrollability of user behavior~\cite{wu2024s}.

Physiological biometrics-based methods focus on the unique physiological features of the ear to distinguish users.
For example, ear images captured using the smartphone camera during a call are employed for authentication~\cite{fahmi2012implicit,damer2012ear,poosarala2018uniform,abate2019implicitly,cherifi2021robust}.
However, these methods are sensitive to environmental conditions, such as low light intensity.
Additionally, active user cooperation is often required to obtain a clear and complete image of the ear.
The smartphone touchscreen can also serve as a capacitive sensor to capture a user's earprint~\cite{Holz2015, rilvan2016user, cabra2022earprint}. 
However, they require the user to active position their ear tightly and fully on the smartphone screen to capture capacitive readings, changing user's call receiving habits. 
Moreover, these methods necessitate rooting the smartphone and modifying the touchscreen module in the kernel source.
Additionally, some methods utilize acoustic signals to sense the ear~\cite{itani2022multimodal, akkermans2005acoustic}. 
However, these methods rely on measuring the ear’s transfer function for authentication, which is highly sensitive to the smartphone’s position. 
As a result, they encounter substantial challenges in maintaining accuracy when the smartphone’s position varies, limiting their practical applicability in real-world scenarios.
Recent studies have explored the use of earphones to assist in authentication on smartphones. 
However, these methods necessitate hardware modifications to existing earphones and the integration of additional sensors, such as cameras~\cite{mizuho2024earauthcam} or inward-facing microphones~\cite{Gao2019, wu2023earae, arakawa2016fast, mahto2018ear, sun2023earmonitor}, which increases costs and leads to incompatibility with commercial earphones.
Furthermore, they require users to constantly carry earphones, significantly compromising convenience and practicality.

Our method does not require any additional hardware or impose extra burden.
It demonstrates resilience to changes in smartphone position and remains effective under various environmental conditions.

\begin{figure*}[!t]
	\centering
	\subfloat[The same user]{\includegraphics[width = 1.68in]{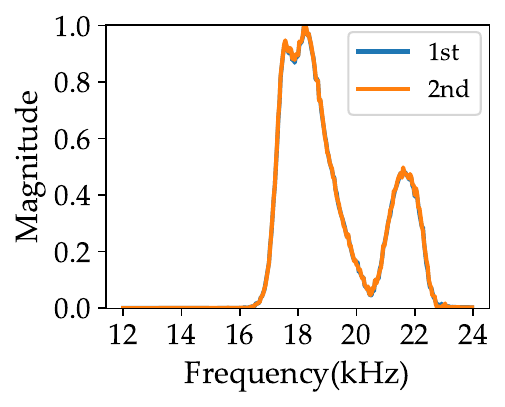}
		\label{one_psd}}
	\hspace{0.1em}
	\subfloat[Two users]{\includegraphics[width = 1.68in]{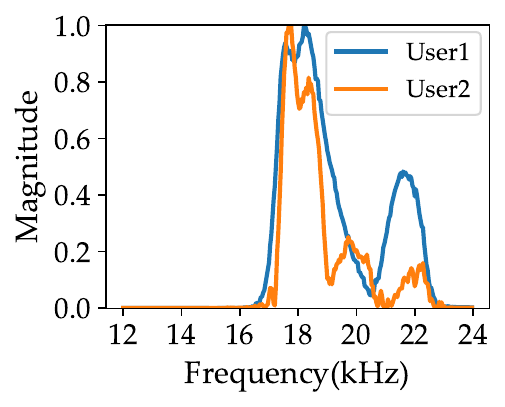}
		\label{two_psd}}
	\hspace{0.1em}
	\subfloat[The same user]{\includegraphics[width = 1.68in]{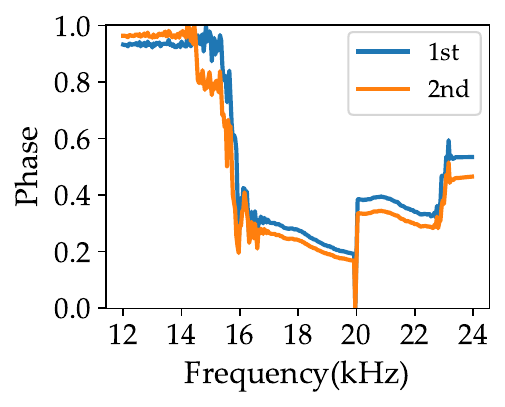}
		\label{one_phase}}
	\hspace{0.1em}
	\subfloat[Two users]{\includegraphics[width = 1.68in]{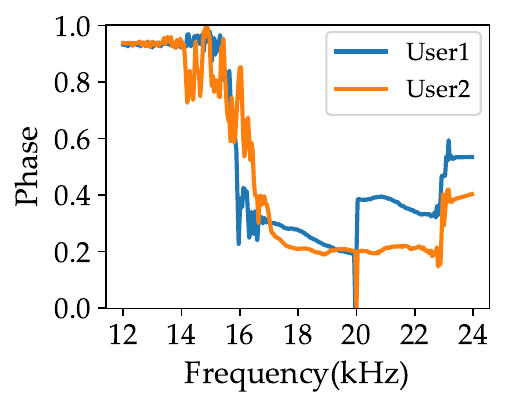}
		\label{two_phase}}
	%	\caption{The magnitude and phase spectrums for two users' signals.}
	\caption{The acoustic profiles of outer ear echoes for two users. (a) Magnitude spectrums for the same user at two times. (b) Magnitude spectrums for two users. (c) Phase spectrums for the same user at two times. (d) Phase spectrums for two users.}
	\label{fig:psdandphase}
\end{figure*}

\begin{figure}
	\begin{minipage}[t]{0.49\linewidth}
		\centering
		\includegraphics[width = 0.9\linewidth]{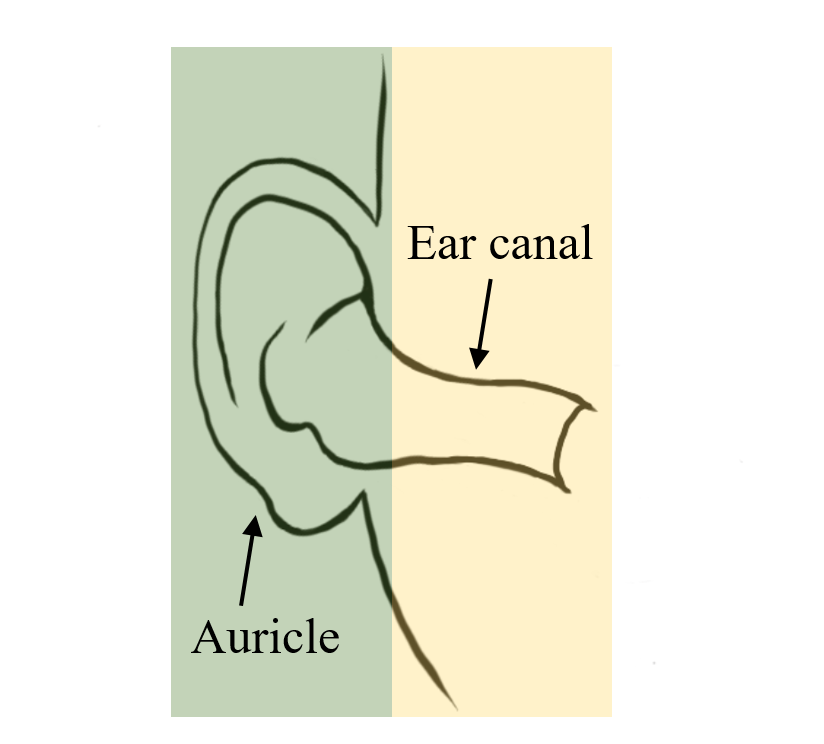}
		%		\caption{Feature separation of four different users.}
		\caption{The structure of the outer ear: auricle and ear canal.}
		\label{fig:outer_ear}
	\end{minipage}
	\hspace{0.01em}
	\begin{minipage}[t]{0.49\linewidth}
		\centering
		\includegraphics[width = 0.9\linewidth]{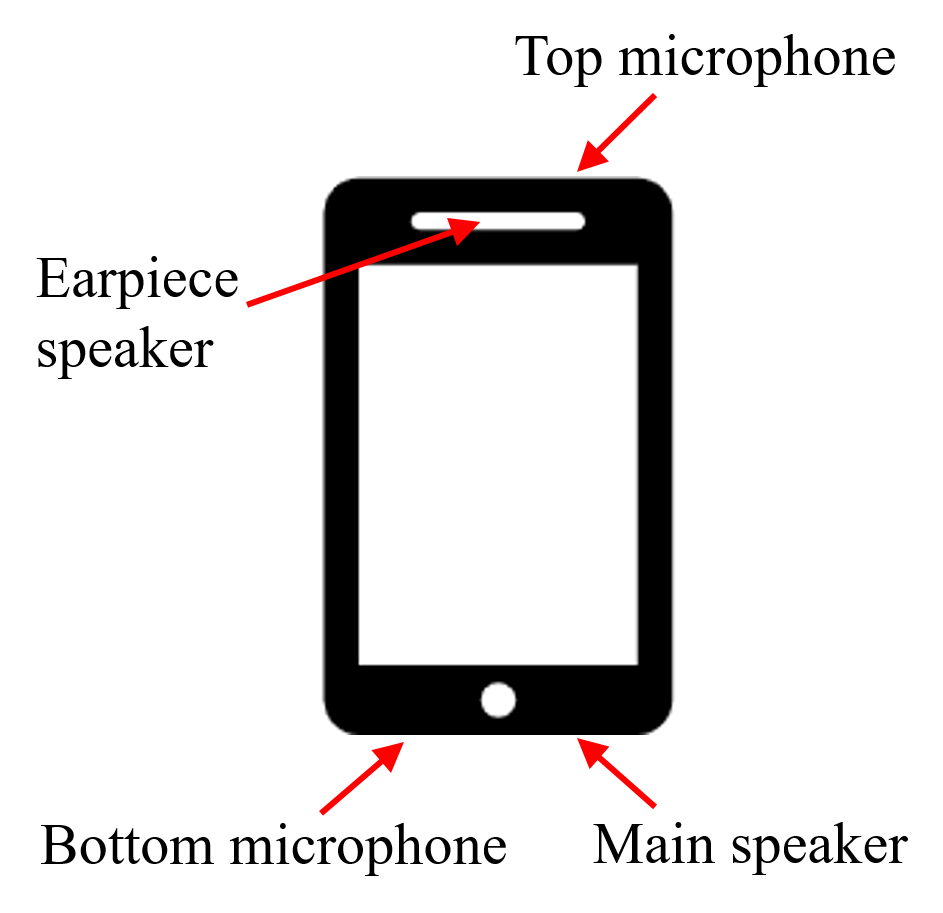}
		\caption{The typical layout of speakers and microphones on smartphones.}
		\label{fig:sensor_layout}
	\end{minipage}
\end{figure}

\subsection{Acoustic Sensing}
Acoustic sensing has garnered significant attention in recent years and finds applications across diverse domains.
Leveraging the capabilities of speakers and microphones, it enables environment sensing~\cite{ren2021proximity,cai2021active,su2024embracing}, the monitoring of human activities such as hand tracking~\cite{chen2017echotrack,wang2022amaging,cheng2022pd}, lip reading~\cite{lu2018lippass,wu2019lvid,zhang2023echospeech}, and breathing monitoring~\cite{nandakumar2015contactless,song2020spirosonic}.
Additionally, acoustic sensing has demonstrated potential in identifying human physiological biometrics, such as hands~\cite{zhou2022presspin,wu2022echohand,yang2023biocase} and faces~\cite{zhou2021robust,wang2023facer,xu2024aface}.

For instance, Cai \emph{et al.}~\cite{cai2021active} employ dual microphones to estimate the speed of air-borne sound propagation, allowing for the inference of ambient temperature.
Echotrack~\cite{chen2017echotrack} determines the distance from the hand to the speaker, enabling continuous hand tracking using triangular geometry.
Lu \emph{et al.}~\cite{lu2018lippass} extract distinctive behavioral features of users’ speaking lips through acoustic signals.
EchoHand~\cite{wu2022echohand} complements camera-based hand geometry recognition by integrating active acoustic sensing for the other hand.
EchoPrint~\cite{zhou2021robust} fortifies face authentication against presentation attacks by emitting inaudible acoustic signals to capture 3D facial features.

Our work uses the inaudible acoustic signal to sense the outer ear without interfering with the normal voice conversation.
Moreover, it provides implicit protection before the call is answered and supports continuous authentication.

\section{Preliminaries}
\label{sec:preliminaries}

\subsection{Outer Ear Echoes}
The outer ear, as the external part of the auditory system, serves as the primary interface between the human body and the acoustic environment.
As depicted in Fig.\ref{fig:outer_ear}, it consists of two main components: the auricle and the ear canal. 
The auricle is a three-dimensional structure formed by cartilage and skin, which is unique in shape and size for every individual\cite{abaza2013survey}. 
The ear canal, a short tube leading to the eardrum, exhibits variations in interspace, curvature, and material composition across populations~\cite{wang2021eardynamic}.

These structural and tissue characteristics of the outer ear significantly influence how acoustic sensing signals are absorbed, reflected, and propagated, producing distinctive echo patterns. 
Specifically, when inaudible acoustic signals are transmitted through the earpiece speaker to actively sense the call receiver's outer ear, they propagate along multiple paths.
During their interaction with the outer ear, part of the signals are absorbed, while others are reflected along different paths.
For instance, the different energy absorption capacities of cartilage and skin subtly modulate signal strength, while the complex geometry of the ear canal affects propagation delays.
As a result, the acoustic signals arrive at the microphone with varying strengths and delays, manifesting as changes in magnitude and phase that encode unique echo patterns.
In this study, we analyze the characteristics of outer ear echoes for authentication.

\subsection{Motivating Examples}
We present a toy example to explore the feasibility of distinguishing between different call receivers based on outer ear echoes.
Two users are employed to simulate the call answering process. 
The Google Pixel 3a is selected as the authentication device, and acoustic data is collected at a sampling rate of 48 kHz.

Specifically, we utilize the earpiece speaker to emit inaudible sensing signals and analyze the resulting echoes from the outer ear using the microphone.
The sensing signal is a 25-millisecond chirp, ranging from 17 kHz to 23 kHz.
After deriving the ear-related signals, we compute their magnitude and phase spectrums.
The results are shown in Fig.~\ref{fig:psdandphase}.
Fig.~\ref{fig:psdandphase}\subref{one_psd} and Fig.~\ref{fig:psdandphase}\subref{one_phase} present the magnitude and phase spectrums for two instances of the same user, respectively.
Fig.~\ref{fig:psdandphase}\subref{two_psd} and Fig.~\ref{fig:psdandphase}\subref{two_phase} show the magnitude and phase spectrums for user 1 and user 2, respectively.
We observe that the profiles of two instances for the same user match each other closely.
In contrast, the profiles for the two users differ in both magnitude and phase. 
These results demonstrate the feasibility of using outer ear echoes for authentication, motivating the design of SCR-Auth.

\subsection{Speaker and Microphone Selection}
Fig.\ref{fig:sensor_layout} illustrates the typical layout of speakers and microphones on modern commercial smartphones. 
These devices are generally equipped with two speakers: a main speaker positioned at the bottom and an earpiece speaker located near the ear\cite{wang2024uface}. 
They also include two microphones: one at the bottom and another at the top for noise cancellation~\cite{yang2020echolock}. 
For our system, we select the earpiece speaker and the top microphone for sending and receiving signals, as their proximity to the ear supports better sensing.

\begin{figure*}[!t]
	\centering
	\includegraphics[width = 0.9\linewidth]{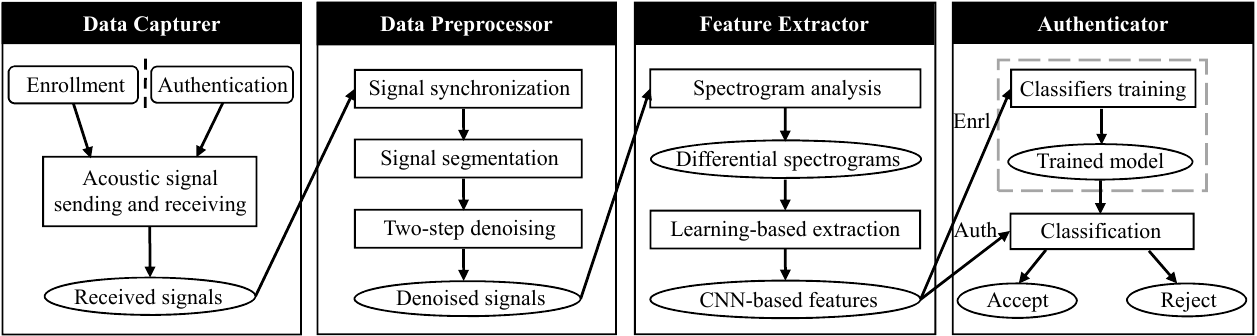}
	\caption{Workflow of SCR-Auth.}
	\label{fig:workflow}
\end{figure*}

\section{Overview of SCR-Auth}
\label{sec:overview}
In this section, we first present the overview of SCR-Auth. Then we introduce the threat model and design goals.

\subsection{System Overview}
The basic idea of SCR-Auth is to utilize the speaker and microphone on a smartphone for outer ear acoustic sensing, and then analyze outer ear biometric features from the received echo signals to authenticate the call receiver.
It consists of two phases: enrollment and authentication.
In the enrollment phase, SCR-Auth builds the authentication model of the legitimate user.
In the authentication phase, SCR-Auth use the built model to determine whether the call receiver is legitimate.

Fig.~\ref{fig:workflow} illustrates the workflow of SCR-Auth, consisting of four key modules: the data capturer, data preprocessor, feature extractor, and authenticator.
The data capturer utilizes the smartphone's earpiece speaker and top microphone as an active sonar system.
It sends inaudible chirp signals and captures the resulting echoes.
The data preprocessor first synchronizes and segments the echo signals through a correlation-based approach.
A two-step denoising process is subsequently applied, which involves the use of a bandpass filter followed by the Magnitude-Phase Spectrogram Subtraction (MPSS) method. 
This approach eliminates ambient noises and other interferences, thereby enhancing the signal from the outer ear.
The feature extractor first performs spectrogram analysis to obtain normalized differential spectrograms.
Then it extracts the reliable features using a pre-trained CNN model based on transfer learning.
The authenticator trains a one-class classification model during the enrollment phase based on the collected samples from the legitimate user.
After enrollment, the model determines whether the user is legitimate.

\subsection{Threat Model}
For the sake of privacy, call receivers typically use the earpiece mode on smartphones to answer calls, holding the smartphone to their ear and listening through the earpiece speaker~\cite{wang2022mmeve}. 
In this paper, we focus on this realistic call-answering scenario.
We assume that the attacker has physical access to the victim’s smartphone when the call comes in.
The attacker’s goal is to bypass the call receiver authentication system, thereby answering the call and performing sensitive operations.
We consider two common types of attacks, based on the attacker’s capabilities and objectives:
\begin{itemize}
	\item \emph{Zero-effort attack.} The attacker has no prior knowledge of the legitimate call receiver and attempts to bypass the authentication system using his/her own ear.
	\item \emph{Mimicry attack.} The attacker observes the legitimate call receiver’s authentication process and then replicates the smartphone's placement near the ear.
\end{itemize}

\subsection{Design Goals}
We think a suitable authentication scheme for a call receiver should satisfy the following goals:
\begin{itemize}
	\item \emph{Accurate and secure:} 
	The scheme should reliably authenticate the legitimate user with a high success rate while accurately rejecting unauthorized users. 
	It should also defend against common attacks.
	\item \emph{Implicit:} 
	The authentication process should not impose additional burden and interfere with normal voice conversations.
	\item \emph{Universal:} 
	It should work on standard commodity smartphones, without requiring additional hardware or root privileges, making it scalable for widespread deployment.
	\item \emph{Robust:} 
	The scheme should be resilient across varying conditions, such as ambient noises, different postures, different periods, and devices.
\end{itemize}

\section{Design of SCR-Auth}
\label{sec:design}

SCR-Auth consists of four modules: data capturer, data preprocessor, feature extractor, authenticator. 
In this section, we provide a detailed explanation of each module.

\subsection{Data capturer}
\label{sec:capture}

\label{sec:sensing}
\begin{figure}[!t]
	\centering
	\subfloat[A chirp signal]{\includegraphics[width = 0.48\linewidth]{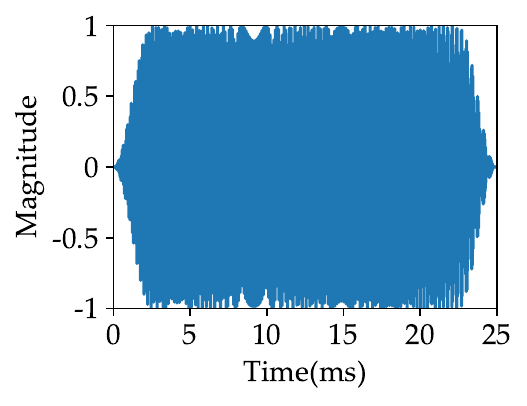}
		\label{fig:send_time}}
	%	\hspace{0.001em}
	\subfloat[The spectrogram]{\includegraphics[width = 0.4568\linewidth]{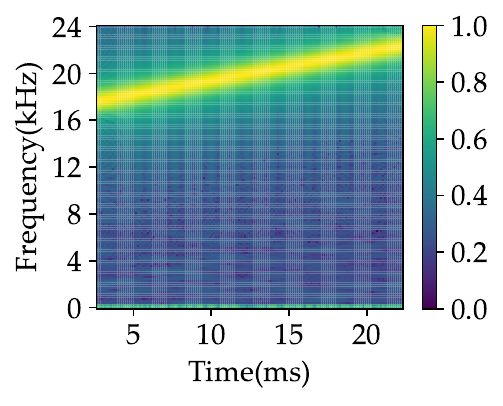}
		\label{fig:send_time_frequency}}
	\caption{Illustration of the designed chirp signal in the time and frequency domains.}
	\label{fig:send}
\end{figure}

SCR-Auth leverages the smartphone's earpiece speaker to emit acoustic sensing signals and the top microphone to receive corresponding echoes.
The data capture process integrates seamlessly with the natural call-receiving procedure. 
Specifically, when a call comes in, the user presses the accept button to answer, which serves as the trigger for the system. 
Upon this action, the earpiece speaker begins emitting inaudible sensing signals, while the top microphone continuously records the resulting echoes for further processing.

SCR-Auth employs chirp signals as acoustic sensing signals, characterized by a continuously varying frequency over time. 
Chirp signals are well-suited for acoustic sensing applications due to their excellent auto-correlation properties~\cite{cai2022ubiquitous}.
Fig.\ref{fig:send} illustrates a designed chirp signal used in this study. 
Research indicates that the upper limit of the human hearing range for adults typically lies between 15–17 kHz\cite{2001Neuroscience}. 
Most smartphones support a maximum sampling rate of 48 kHz~\cite{wu2022echohand}, which limits the sensing signal's maximum frequency to below 24 kHz in compliance with the Nyquist sampling theorem~\cite{ba2020learning}.
%To ensure a broad sensing range while minimizing auditory discomfort, 
To ensure a broad sensing range while remaining imperceptible to users, we adopt the 25-millisecond chirp signal sweeping from 17 kHz to 23 kHz, a range commonly used in acoustic sensing applications~\cite{wang2021eardynamic}. 
The first and last 120 samples of the chirp are tapered using a Hamming window to reduce potential acoustic annoyance~\cite{ifeachor2002digital}. 
The interval between two chirps is set to 25 milliseconds, resulting in a sensing signal that alternates between a 1200-sample chirp and a 1200-sample silent period.

\begin{figure}[!t]
	\centering
	\subfloat[The transmitted pilot signal]{\includegraphics[width = 0.4855\linewidth]{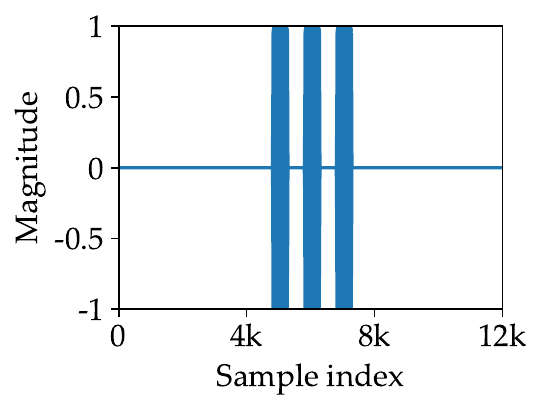}
		\label{fig:pilot_send}}
	%	\hspace{0.0001em}
	\subfloat[The received pilot signal]{\includegraphics[width = 0.4795\linewidth]{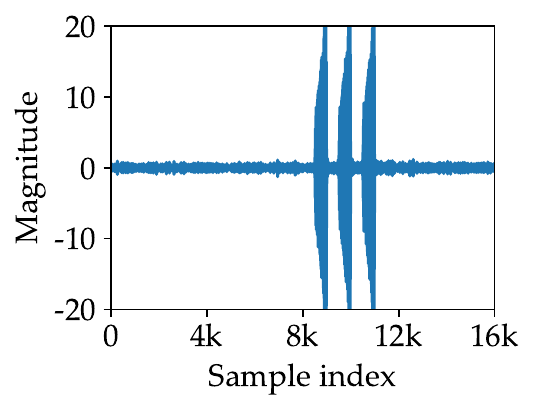}
		\label{fig:pilot_receive}}
	\caption{The pilot signal for synchronizing the smartphone speaker and microphone.}
	\label{fig:pilot}
\end{figure}
\begin{figure}[!t]
	\centering
	\includegraphics[width = 0.78\linewidth]{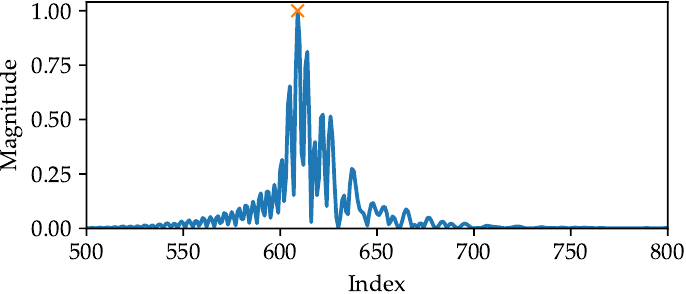}
	\caption{An example of the cross-correlation result.}
	\label{fig:corr}
\end{figure}

\begin{figure*}[!t]
	\centering
	\subfloat[User 1]{\includegraphics[width = 1.66in]{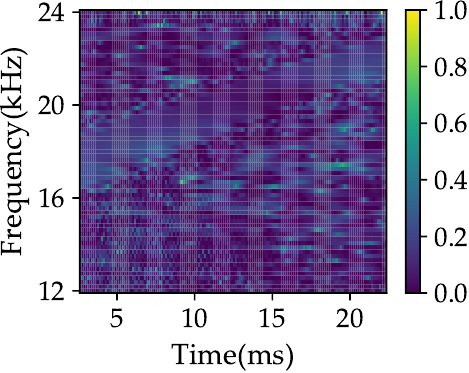}
		\label{fig:fea_m1}}
	\hspace{0.25em}
	\subfloat[User 2]{\includegraphics[width = 1.66in]{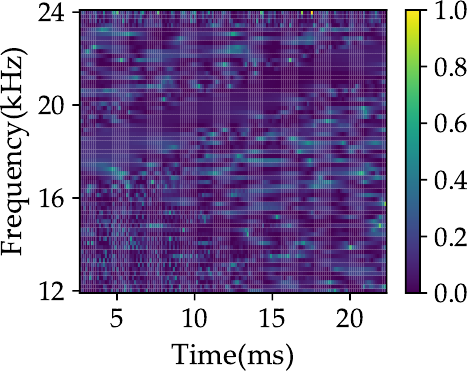}
		\label{fig:fea_m3}}
	\hspace{0.25em}
	\subfloat[User 1]{\includegraphics[width = 1.66in]{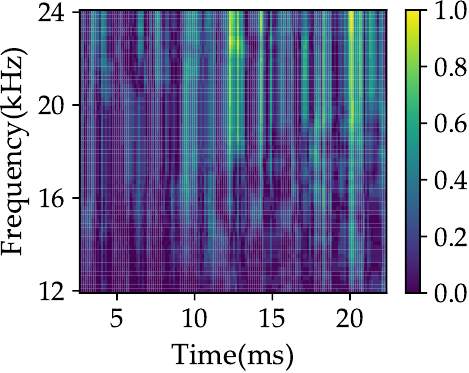}
		\label{fig:fea_p1}}
	\hspace{0.25em}
	\subfloat[User 2]{\includegraphics[width = 1.66in]{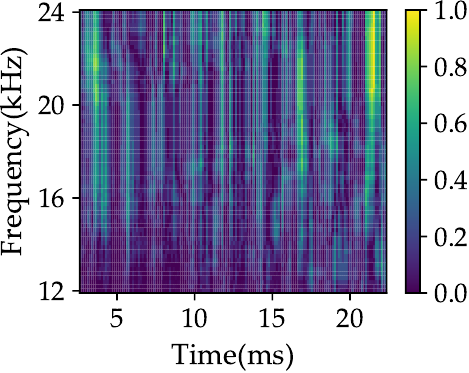}
		\label{fig:fea_p3}}
	\caption{Normalized magnitude and phase spectrograms for two users. (a) A magnitude spectrogram for user 1. (b) A magnitude spectrogram for user 2. (c) A phase spectrogram for user 1. (d) A phase spectrogram for user 2.}
	\label{fig:feature}
\end{figure*}

\subsection{Data Preprocessor}
\label{sec:preprocess}
After capturing the acoustic signals, we proceed with a series of preprocessing steps: synchronization, segmentation, and denoising.

\textbf{Signal synchronization and segmentation.}
To ensure precise segmentation of the acoustic signals, we propose a two-step synchronization approach that aids in the alignment of the signals for further analysis.

Initially, a pilot signal is appended before the sensing sequence to provide coarse synchronization between the smartphone's speaker and microphone~\cite{tung2016expansion}. 
This pilot signal consists of three 500-sample chirps, sweeping from 22 kHz to 18 kHz. 
An example of the transmitted pilot signal is shown in Fig.\ref{fig:pilot}\subref{fig:pilot_send}, with the corresponding received pilot signal depicted in Fig.\ref{fig:pilot}\subref{fig:pilot_receive}. 
By detecting the presence of this pilot, we can identify the starting point of the sensing process during the call reception.
Once coarse synchronization is achieved, the system proceeds to divide the received signals into 50-millisecond segments, each corresponding to a single sensing event. 

In the second step, a finer level of synchronization is applied within each segment to counteract any timing drifts or distortions caused by the transmission channel. 
For each segment, a matched filter is used to precisely determine the arrival time of the transmitted chirp signal~\cite{chen2020chaperone}. 
Specifically, the cross-correlation $R_{xy}$ between a received signal segment $y(t)$ and the transmitted chirp signal $x(t)$ is calculated, as expressed in Eq.~\ref{eq:correlation}.
\begin{equation} 
	\label{eq:correlation} 
	R_{xy} = y(t) \ast x^{\star}(-t) 
\end{equation}
Here, $\ast$ denotes the convolution operator, and $x^{\star}(-t)$ is the complex conjugate of $x(-t)$. 
Fig.~\ref{fig:corr} illustrates an example of the cross-correlation result.
The index of the highest peak of the cross-correlation result is identified as the start point.
Based on the length of the chirp signal, we finally derive 1200-sample segments.

\textbf{Signal denoising.}
Due to the multipath effect, the received signals include not only outer ear echoes, but also direct path signals and environmental reflections. 
Additionally, ambient noises are inevitably introduced during sound propagation.
In this study, we propose a two-step denoising approach that combines bandpass filtering with magnitude-phase spectrogram subtraction (MPSS) to effectively suppress unwanted interference.

In the first step, we address ambient noises by applying a Butterworth bandpass filter to remove out-of-band interference~\cite{xie2022teethpass}. 
The filter’s cutoff frequencies are set at 17 kHz and 23 kHz, corresponding to the expected frequency range of the chirp signal. 
This selective filtering ensures that only the relevant frequency components are retained, thereby improving the signal-to-noise ratio.

In the second step, we apply the MPSS method to suppress the interference from direct path signals and environmental reflections.
The key idea is to carefully choose a reference segment that primarily contains direct path signals and environmental reflections, devoid of outer ear echoes.
Since interference components, such as direct path signals and static objects reflections, remain consistent during sensing.
By subtracting these interference components, we can highlight echoes from the outer ear.

By analyzing the process of call reception, we select the the first signal segment, captured immediately after the user clicks the "accept" button, as the reference segment. 
At this point, the smartphone is typically stationary and has not yet been placed on the ear.
Once the smartphone is positioned on the ear, changes in the received signal can be attributed to echoes from the ear. 
The reference segment plays two crucial roles: it acts as a template for the direct path signal, eliminating the need for a quiet environment to detect this signal, and provides a baseline for environmental interference during the call.

To perform MPSS, we use the Short-Time Fourier Transform (STFT)~\cite{chen2023fall} to compute the magnitude and phase spectrograms for each signal segment, then construct differential spectrograms based on the selected reference segment.
Denoting the magnitude spectrogram as $\bm{S}_m$ and the phase spectrogram as $\bm{S}_p$, the combined magnitude-phase spectrograms can be expressed as $\bm{Spec} = [\bm{S}_m;\bm{S}_p]$.
The differential spectrogram, represented as $\triangle \bm{Spec} = [\triangle \bm{S}_m;\triangle \bm{S}_p]$, is then calculated according to the Eq.~\ref{eq:abs}:
\begin{equation}
	\label{eq:abs}
%	\triangle \bm{Spec} = |\bm{Spec}_{s} - \bm{Spec}_{r}|
	\triangle \bm{Spec} = |\bm{Spec}_{s} - \bm{Spec}_{r}| 
\end{equation}
where  $\bm{Spec}_{r}$ represents the spectrograms of the reference segment, and $\bm{Spec}_{s}$ corresponds to the spectrograms of one sensing segment.
The differential spectrogram serves as the foundation for feature extraction.

\subsection{Feature Extractor}
\label{sec:feature}
In this section, we perform spectrogram analysis and use a pre-trained convolution neural network model to extract reliable features.

\textbf{Spectrogram analysis.}
The acoustic signals captured by the top microphone interact with the user's outer ear, undergoing absorption and reflection, leading to variations in both signal strength and time delay. 
Consequently, we focus on magnitude and phase spectrograms to represent these variations, as they contain valuable biometric information from the outer ear.

To process the differential spectrogram $\triangle \bm{Spec}$ extracted by the preprocessor, we first reduce computation overhead by focusing on specific components. 
Then we apply min-max normalization~\cite{liu2022secure} to scale the spectrogram values to the range of [0, 1].
We retain signal components with frequencies above a threshold $f_{thre}$, which is empirically set to 12 kHz.
The refined differential spectrogram is represented as $\triangle \bm{Spec}_{emp} = [\triangle \bm{S}_{mr};\triangle \bm{S}_{pr}]$, where $\triangle \bm{S}_{mr} = \triangle \bm{S}_m(I_{thre}:, :)$ and $\triangle \bm{S}_{pr} = \triangle \bm{S}_p(I_{thre}:, :)$.
$I_{thre}$ represents the FFT bin index corresponding to the threshold frequency $f_{thre}$.
Given the sampling rate of $f_{s} = 48kHz$ and FFT points $N_{fft} = 256$, the $I_{thre}$ is calculated as $\frac{f_{thre}\times N_{fft}}{f_{s}} = 64$.
This results in a spectrogram with dimensions $65\times158\times2$.
Finally, the normalized spectrogram $\triangle \bm{Spec}_{norm}$ is computed using Eq.~\ref{eq:norm}.
\begin{equation}
	\label{eq:norm}
	\triangle \bm{Spec}_{norm} = \frac{\triangle \bm{Spec}_{emp} - min(\triangle \bm{Spec}_{emp})}{max(\triangle \bm{Spec}_{emp}) - min(\triangle \bm{Spec}_{emp})}
\end{equation}

As an example, we present the normalized magnitude and phase spectrograms of two users in Fig.~\ref{fig:feature}.
We can observe that spectrograms show differences for different users.
These spectrograms are later used as inputs for model training.

\begin{table}[!t]
	\centering
	\caption{The structure of our base CNN model.}
	\label{tab:cnn}
	\renewcommand\arraystretch{1.01}
	\begin{tabular}{llll}
		\hline
		Layer & Layer type      & Output shape & \# Param \\
		\hline
		1     & Conv2D + ReLU   & (63,156,16)  & 304      \\
		2     & Conv2D + ReLU   & (61,154,16)  & 2,320    \\
		3     & Max Pooling     & (30,77,16)   & 0        \\
		4     & Dropout         & (30,77,16)   & 0        \\
		5     & Conv2D + ReLU   & (28,75,32)   & 4,640    \\
		6     & Conv2D + ReLU   & (26,73,32)   & 9248     \\
		7     & Max Pooling     & (13,36,32)   & 0        \\
		8     & Dropout         & (13,36,32)   & 0        \\
		9     & Conv2D + ReLU   & (11,34,16)   & 4,624    \\
		10    & Conv2D + ReLU   & (9,32,16)    & 2,320    \\
		11    & Max Pooling     & (4,16,16)    & 0        \\
		12    & Dropout         & (4,16,16)    & 0        \\
		13    & Flatten         & (1024)       & 0        \\
		14    & Dense + ReLU    & (128)        & 131,200  \\
		15    & Dropout         & (128)        & 0        \\
		16    & Dense + Softmax & (30)         & 3,870    \\
		\hline
	\end{tabular}
\end{table}

\textbf{Learning-based feature extraction.}
To extract reliable features from magnitude-phase spectrograms, we design a learning-based feature extractor to mitigate the variability caused by smartphone position changes. 
The foundation of this extractor is a convolutional neural network (CNN) with superior capabilities in feature extraction and representation~\cite{boutros2022pocketnet,boutros2023exfacegan}. 
Leveraging multi-user data collected under diverse natural smartphone positions during call reception, we train the CNN model using supervised learning to extract identity-related features while disregarding secondary factors, such as changes in the relative position between the ear and smartphone.
Based on transfer learning~\cite{wu2020liveness}, we remove the final layer of the pre-trained CNN and use the output from the 15th layer (as detailed in Table~\ref{tab:cnn}) as a generalized feature extractor. 
This approach enables the network to effectively capture effective features of the outer ear.

Table~\ref{tab:cnn} presents the architecture of our base CNN model, which is designed with multiple convolutional layers to effectively extract features.
Each two-dimensional convolutional (Conv2D) layer employs the rectified linear unit (ReLU) as its activation function, mitigating the vanishing gradient problem.
The max-pooling layer is used to down-sample the data from the previous activation layer, which reduces the data dimension and saves computational costs.
Dropout layers are added after the max pooling layers to prevent overfitting.
The final layer of the model is a dense layer with a softmax activation function, which outputs the probability distribution for each class. 
The kernel sizes for the Conv2D and max pooling layers are set to $3\times3$ and $2\times2$, respectively.
The whole model contains 158,526 parameters.

The base CNN model is trained using data from 30 participants, with each contributing 500 acoustic samples.
Aligned with natural call reception habits, participants are asked to place the smartphone down and pick it up again, simulating a variety of smartphone positions.
We employ the Adam optimizer for parameter optimization and use categorical cross-entropy as the loss function.
The training process is performed with a batch size of 50 over 10 epochs.
Once trained, the base model serves as the foundation of our feature extractor, eliminating the need for retraining when applied to unseen users.
Leveraging the concept of transfer learning, we transform the pre-trained base model into a generalized feature extractor by removing its final layer (i.e., the 16th layer) and retaining the preceding layers.
This transformation results in a lightweight 659 kB feature extractor, optimized for deployment on mobile devices.
Finally, the feature extractor generates a 128-dimensional feature vector, which is utilized in each authentication process to ensure efficient performance.

\begin{figure}[!t]
	\centering
	\includegraphics[width=0.48\linewidth]{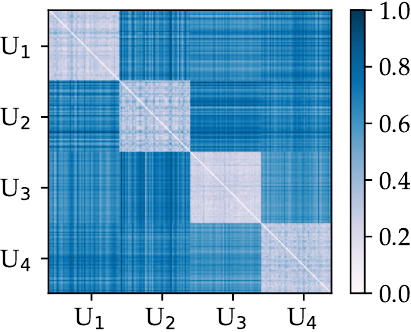}
	\caption{Euclidean distance between the CNN-based features of four users.}
	\label{fig:heatmap}
\end{figure}

To investigate the effectiveness of the CNN-extracted features, we randomly select 200 test samples from four users. 
Using the CNN-based feature extractor, we calculate the Euclidean distance between the features of these samples. 
Fig.~\ref{fig:heatmap} illustrates the results, where the x-axis and y-axis represent the feature points for the four users, and the intersections indicate the normalized Euclidean distances. 
The experimental results reveal that features extracted from the same user exhibit high similarity, while those from different users demonstrate low similarity. 
This highlights the effectiveness of the CNN-based features in distinguishing between users.

\subsection{Authenticator}
\label{sec:classifier}
The training dataset in our scenario exclusively consists of samples from the legitimate call receiver.
Therefore, it can be considered a one-class classification problem, commonly known as a novelty detection problem~\cite{wu2020caiauth}.
We utilize data samples from the legitimate user to train a classifier, employing the feature vectors extracted from the CNN-based model.
Subsequently, we assess whether the call receiver is legitimate based on the classifier's judgment.
We consider two standard novelty detection methods to classify users: one-class support vector machine (OCSVM)~\cite{scholkopf1999support} and local outlier factor (LOF)~\cite{breunig2000lof}.

\section{Data Collection}
\label{sec:data collection}
To collect the experiment data, we develop an Android data collection app.
We use the earpiece speaker to send inaudible sensing signals and the top microphone as the receiver.
After receiving approval from our university's institutional review board (IRB), we started our data collection.
We recruited 37 participants, aged from 20 to 27 (graduate and undergraduate students), including 19 males and 18 females.
We explicitly informed the participants that the purpose of the experiments was to authenticate the receiver of a call.
Similar to answering a call, participants were required to click the start button and picked up the smartphone toward their ear.
They were allowed to make slight adjustments to the smartphone's position to cover different situations. 
In our data collection, we compiled the following 8 datasets.

\textbf{Dataset-1.}
%30*500
This dataset is used to train our CNN-based feature extraction model.
We recruited 30 participants to collect acoustic signals on Google Pixel 3a.
For each of them, we collected 500 acoustic signals.
In total, we collected $30\times500=15,000$ acoustic signals for CNN model training.

\textbf{Dataset-2.}
%30*500
This dataset is utilized to evaluate the overall performance of our system, which is collected under basic settings.
We collected acoustic sensing data from 30 participants on Google Pixel 3a.
Participants were seated naturally in a quiet environment.
We collected 500 acoustic signals for each participant.
Besides, we collected acoustic sensing data from 7 unseen participants to evaluate the performance of the CNN-based feature extraction model for new users.
We collected 500 acoustic signals for each new participant.

\textbf{Dataset-3.}
To evaluate the performance of continuous authentication, we collected acoustic sensing data from two situations: listening and speaking.
Therefore, we recruited 5 participants and performed acoustic sensing every 1s.
We collected 600 acoustic signals for each participant while they were solely listening and another 600 acoustic signals while they were speaking. 
In total, we collected $5 \times 600 \times 2 = 6,000$ acoustic signals for dataset-3.

\textbf{Dataset-4.}
%30*500
To evaluate the influence of ambient noises, we use a laptop as the noise source to simulate the noisy environment.
The laptop played the song ’Human Sound/Restaurant2’ at 50\% volume, which contains common noises in daily life.
The sound pressure in this noise environment is about 60-62dB.
30 participants performed this experiment.
We collected 500 acoustic signals for each participant in the noisy environment.
Dataset-4 involves $30\times500=15,000$ acoustic signals.

\begin{figure*}
	\begin{minipage}[t]{0.32\linewidth}
		\centering
		\includegraphics[width = 0.92\linewidth]{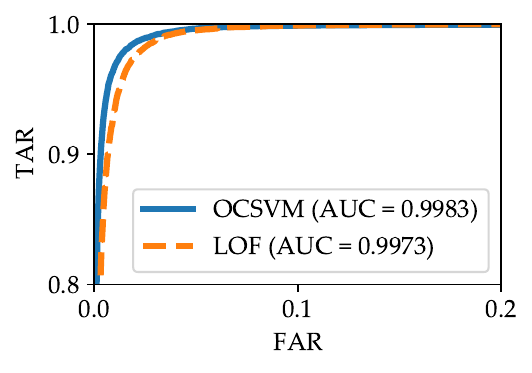}
		\caption{ROC curves of two classifiers with the best parameters.}
		\label{fig:roc_total}
	\end{minipage}
	\hspace{0.5em}
	\begin{minipage}[t]{0.32\linewidth}
		\centering
		\includegraphics[width = 1\linewidth]{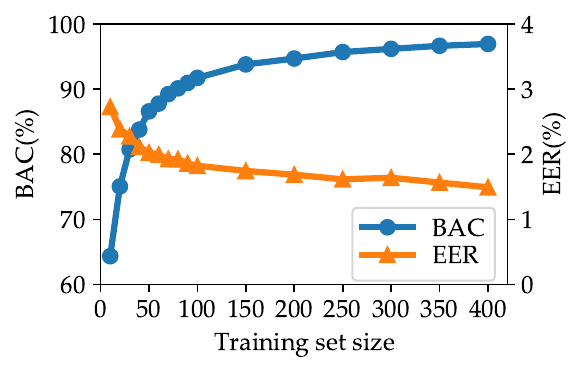}
		\caption{The BAC and EER for different training set sizes.}
		\label{fig:size}
	\end{minipage}
	\hspace{0.5em}
	\begin{minipage}[t]{0.32\linewidth}
		\centering
		\includegraphics[width = 0.91\linewidth]{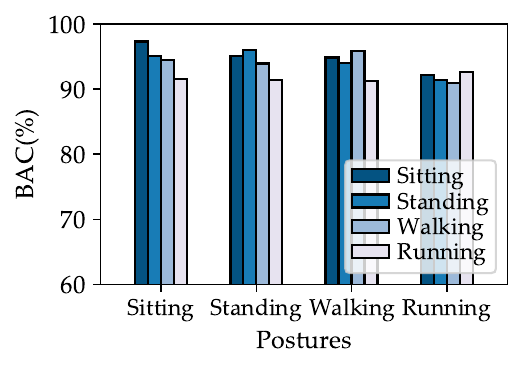}
		\caption{The BAC of SCR-Auth trained and evaluated under different postures.}
		\label{fig:posture}
	\end{minipage}
\end{figure*}

\textbf{Dataset-5.}
%30*500
To evaluate the authentication performance over time, we collected data from different time periods.
Dataset-2 is collected in the first round of collection.
For 30 participants, we collected data one week and two weeks after the first collection round.
For each round of collection, the acoustic signals are $30\times500=15,000$.
We finally got 30,000 acoustic signals for dataset-5.

\textbf{Dataset-6.}
%4*10*250
To evaluate the influence of human postures, we consider four common postures: sitting, standing, walking, and running.
Dataset-2 was collected under the sitting posture.
In this dataset, we recruited 10 participants and collected acoustic data for standing, walking, and running postures.
For each participant, we collected 250 acoustic signals for each posture.
Finally, we obtained $10\times3\times250=7,500$ acoustic signals.

\textbf{Dataset-7.}
%3*10*500
To evaluate the performance of our system on different devices, we collected acoustic data on two extra smartphones: Google Pixel 4 and Vivo S12.
10 participants are recruited to do this experiment.
For each participant, we collected 500 acoustic signals on each device.
As a result, we got $10\times2\times500=10,000$ acoustic signals for dataset-7.

\textbf{Dataset-8.}
To evaluate the system defense against attacks, we chose 7 participants to serve as attackers.
Then we evaluated two types of attacks: zero-effort attack and mimicry attack.
i) Dataset-8A. 
For the zero-effort attack, 7 participants attempted to guess how the legitimate user performs the authentication process.
We finally got $7\times500=3,500$ acoustic signals on Google Pixel 3a for dataset-8A.
ii) Dataset-8B.
For the mimicry attack, the attacker observes and imitates the authentication process of legitimate users.
Specifically, each attacker chose 5 participants to carefully observe and then imitate their authentication process.
We finally got $7\times500=3,500$ acoustic signals on Google Pixel 3a for dataset-8B.

\section{Evaluation}
\label{sec:evaluation}
In this section, we report the evaluation results of the proposed system.
We first present the evaluation metrics, and show the overall performance of SCR-Auth.
Additionally, we evaluate its effectiveness under different settings and security against attacks. 
Finally, we present the authentication latency of our system.

\subsection{Evaluation Metrics}
There are four possible results of classification: True acceptance (TA), True rejection (TR), False acceptance (FA), False rejection (FR).
We use the following metrics to evaluate the performance of SCR-Auth.
True acceptance rate is defined as $TAR = \frac{TA}{TA+FR}$, which measures the proportion of samples classified as positive among legitimate user samples.
True rejection rate is defined as $TRR = \frac{TR}{TR+FA}$, which measures the proportion of samples classified as negative among illegal user samples.
Balanced accuracy (BAC) is the average of true acceptance rate and true rejection rate, which is defined as $BAC = \frac{1}{2}(TAR +TRR)$.
It is used to evaluate the accuracy of imbalanced datasets.
A higher BAC means better performance of the system.
False acceptance rate ($FAR = \frac{FA}{FA+TR}$) represents the rate at which illegal samples are wrongly accepted.
False rejection rate ($FRR = \frac{FR}{FR+TA}$) represents the rate at which legitimate samples are wrongly rejected.
Receiver operation characteristic (ROC) shows dynamic changes of TAR against FAR at different classification thresholds.
The area under the ROC curve (AUC) is used to measure the probability that prediction scores of legitimate users are higher than illegal users.
Equal error rate (EER) is the point on the ROC curve, where FAR is equal to FRR.
A larger AUC and lower EER mean better performance of the system.

\begin{figure*}
	\begin{minipage}[t]{0.32\linewidth}
		\centering
		\includegraphics[width = 0.97\linewidth]{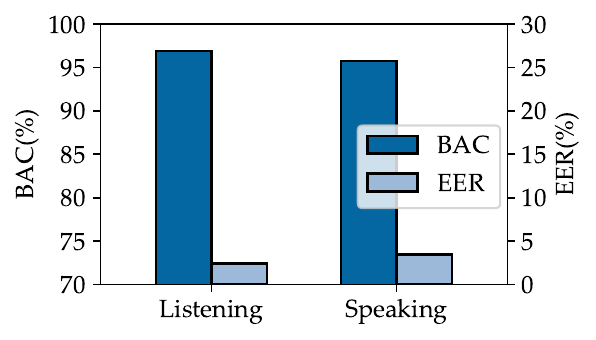}
		\caption{The BAC and EER performance for continuous authentication.}
		\label{fig:continuous}
	\end{minipage}
	\hspace{0.5em}
	\begin{minipage}[t]{0.32\linewidth}
		\centering
		\includegraphics[width = 0.97\linewidth]{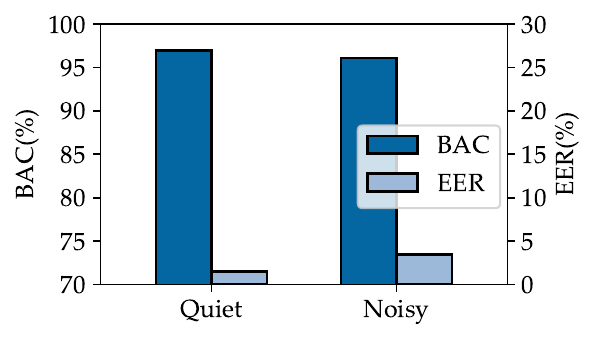}
		\caption{The BAC and EER performance under different noise conditions.}
		\label{fig:noise}
	\end{minipage}
	\hspace{0.5em}
	\begin{minipage}[t]{0.32\linewidth}
		\centering
		\includegraphics[width = 0.97\linewidth]{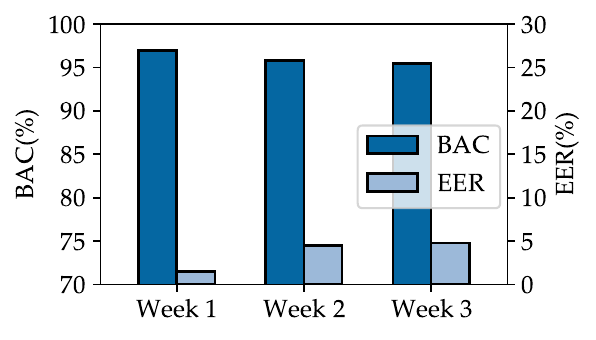}
		\caption{The BAC and EER performance at different time periods.}
		\label{fig:time}
	\end{minipage}
\end{figure*}

\begin{table}[!t]
	\centering
	\caption{Mean/standard deviation of BAC(\%), EER(\%), and AUC under two different one-class classifiers.}
	\label{tab:all}
	\renewcommand\arraystretch{1.01}
	\begin{tabular}{c|ccc}
		\hline
		Classifier & Mean/Std BAC & Mean/Std EER & Mean/Std AUC  \\
		\hline
		OCSVM      & 96.95/1.45   & 1.53/1.35    & 0.9982/0.0025 \\
		LOF        & 96.13/3.18   & 1.90/1.56    & 0.9972/0.0034 \\
		\hline
	\end{tabular}
\end{table}

\begin{table}[!t]
	\centering
	\caption{Mean/standard deviation of BAC(\%), EER(\%), and AUC for new users.}
	\label{tab:newuser}
	\renewcommand\arraystretch{1.01}
	\begin{tabular}{c|ccc}
		\hline
		Classifier & Mean/Std BAC & Mean/Std EER & Mean/Std AUC  \\
		\hline
		OCSVM      & 96.48/1.63   & 2.78/1.95    & 0.9955/0.0043 \\
		LOF        & 93.49/4.59   & 2.89/2.21    & 0.9946/0.0063 \\
		\hline
	\end{tabular}
\end{table}

\subsection{Overall Performance}

\textbf{Performance of different classifiers.}
\label{sec:overall}
We use 30 users in dataset-2 to evaluate the authentication effectiveness of SCR-Auth.
We employ a 5-fold cross-validation for each user to split the data and train a one-class classifier.
Then we test the classifier model using the remaining data of the user as well as data from other users.

This study considers two types of one-class classifiers: one-class support vector machine (OCSVM) and local outlier factor (LOF).
Parameters such as the $kernel$, $\gamma$, and $\nu$ significantly impact the results for OCSVM, while for LOF, we consider the $n\_neighbors$ parameter. 
We employ grid search to find the best parameter combinations for each classifier.
Ultimately, we determine that the radial basis function kernel works best for OCSVM, with $\gamma$ = 'scale' and $\nu$ = 0.01. 
For LOF, the optimal $n\_neighbors$ value is 3.
Fig.~\ref{fig:roc_total} presents the ROC curves of the two classifiers with the best parameters.
The AUC for OCSVM is 0.9983, and for LOF, it is 0.9973.
A higher AUC value suggests better system performance. 
The results indicate that the OCSVM classifier outperforms the LOF classifier.
Table~\ref{tab:all} shows the mean and standard deviation of BAC, EER, and AUC metrics under two classifiers.
OCSVM demonstrates superior BAC and EER metrics compared to LOF, thus we select it as our classifier for subsequent evaluations.
This experiment reveals that SCR-Auth achieves an average BAC of 96.95\% and an EER of 1.53\% using the OCSVM classifier.
These results indicate that SCR-Auth is effective in distinguishing users.

\begin{figure}[!t]
	\centering
	\includegraphics[width=1\linewidth]{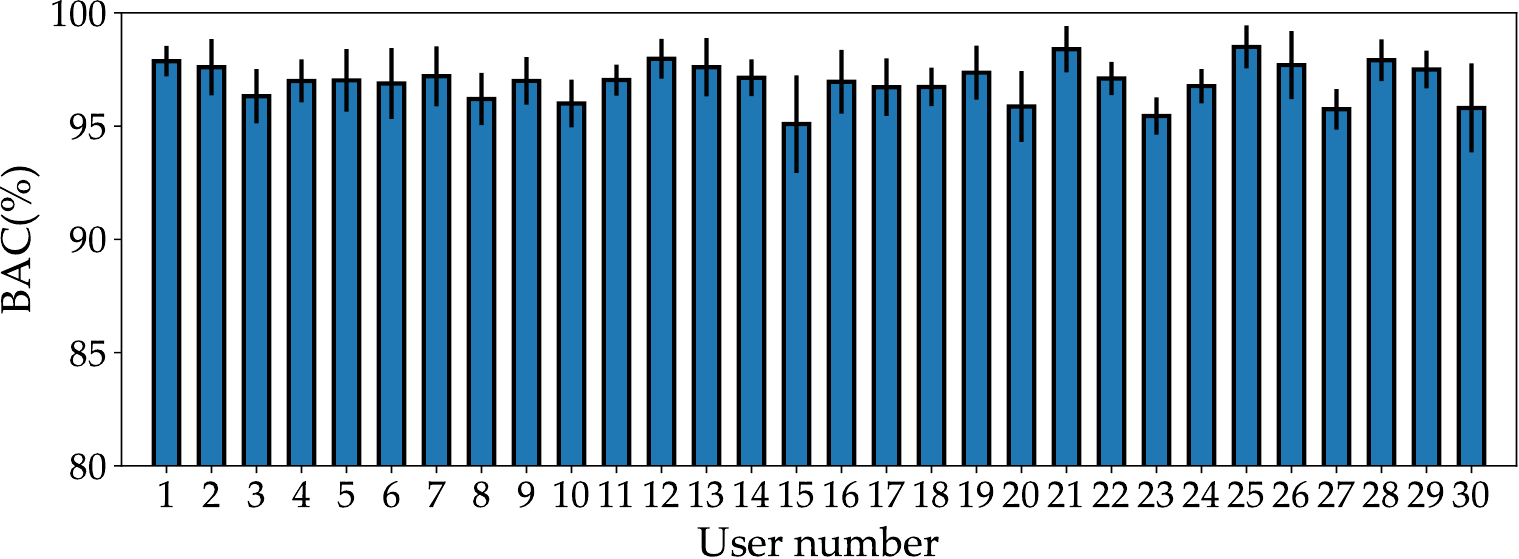}
	\caption{BAC performance for each user.}
	\label{fig:peruser}
\end{figure}

\textbf{Per-user breakdown analysis.}
To evaluate the performance of SCR-Auth across 30 different users, we present the BAC of each user under the OCSVM classifier, as shown in Fig.~\ref{fig:peruser}.
Notably, user \#25 achieves the highest BAC of 98.5\%, marking the best case among all participants. 
While the performance of SCR-Auth varies across users, the BAC for every user exceeds 95\%, demonstrating the overall effectiveness of SCR-Auth.

\textbf{Performance of feature extractor on unseen users.}
To evaluate the performance of the CNN-based feature extractor on new users, we use data from 7 unseen participants, as described in dataset-2, who are not included in the CNN model's pre-training.
We use 5-fold cross-validation to split the data.
Then we train a one-class SVM (OCSVM) classifier and a local outlier factor (LOF) classifier for each participant.
Table~\ref{tab:newuser} shows the mean and standard deviation of BAC, EER, and AUC metrics for new users under two classifiers.
The BACs for OCSVM and LOF are 96.48\% and 93.49\%, respectively.
Compared to results in Table~\ref{tab:all}, the BAC falls 0.47\% for OCSVM and falls 2.64\% for LOF.
For the OCSVM classifier, the BAC is over 96\%, demonstrating the feature extractor's effectiveness for new users.
Although the feature extractor is trained on limited data, it is still available to a wide range of users.

\textbf{Performance of continuous authentication.}
We analyze two common situations to evaluate the performance of continuous authentication.
During the process of answering a call, the receiver will be in one of two states: listening to the caller or speaking to the caller.
We train on dataset-2 and test on dataset-3 for evaluation.
The results are shown in Fig.~\ref{fig:continuous}.
For listening and speaking states, the BACs are 96.89\% and 95.73\%, respectively.
The EERs are 2.39\% and 3.46\%, respectively.
The experimental results show that SCR-Auth is available for continuous authentication.

\subsection{Impact Factors Study}

\textbf{Impact of ambient noises.}
To assess the impact of ambient noise on system performance, we compare the results under different noise conditions. 
In this experiment, dataset-2, which is collected in a quiet environment, is used for training. 
We then evaluate the system's performance on both dataset-2 (for quiet conditions) and dataset-4 (for noisy conditions). 
Fig.~\ref{fig:noise} shows the BACs and EERs in both quiet and noisy environments. The BACs are 96.95\% and 96.09\%, and the EERs are 1.53\% and 3.45\%, respectively.
These results present that SCR-Auth is available for different noise conditions.

\begin{table}[!t]
	\centering
	\caption{Mean/standard deviation of BAC(\%), EER(\%), and AUC for three different devices.}
	\label{tab:device}
		\renewcommand\arraystretch{1.01}
	\begin{tabular}{c|ccc}
		\hline
		Classifier & Mean/Std BAC & Mean/Std EER & Mean/Std AUC  \\
		\hline
		Pixel 3a   & 97.32/1.55   & 0.85/0.80    & 0.9994/0.0011 \\
		Pixel 4    & 97.62/1.47   & 0.86/1.29    & 0.9989/0.0027 \\
		Vivo S12   & 95.20/1.19   & 4.03/1.02    & 0.9926/0.0036 \\
		\hline
	\end{tabular}
\end{table}

\begin{table}[!t]
	\centering
	\caption{Bypassed samples, FAR(\%) and mean prediction scores under two different attacks.}
	\label{tab:attack}
		\renewcommand\arraystretch{1.01}
	\begin{tabular}{c|ccc}
		\hline
		Attack             & Bypassed samples & FAR  & Prediction scores \\
		\hline
		Zero-effort attack & 33(3500)         & 0.94 & -0.39             \\
		Mimicry attack     & 49(3500)         & 1.40 & -0.42             \\
		\hline
	\end{tabular}
\end{table}

\textbf{Impact of training dataset size.}
To investigate the impact of training set size, we change the amount of training data points for each user on dataset-2.
Specifically, for each user, we vary the training data points from 10 to 400 in steps of 10 or 50 to train a one-class SVM classifier.
Then we test on the rest of the data.
Fig.~\ref{fig:size} shows the BAC and EER for different training set sizes.
As the size of the training set increases from 10 to 400, the BAC rises from 64.35\% to 96.94\%.
The EER falls to 1.49\% from 2.73\% when the training set size increases from 10 to 400.
That may be because the classifier can learn a better boundary with more legitimate data.
The BAC is over 90\% with 80 training data points and is over 95\% with 200 training data points.
With 50 training data points, the EER is less than 2\%.
These results show that our system is practical on mobile devices.

\textbf{Impact of different postures.}
To evaluate the impact of different postures, we use 10 participants' data in dataset-2 and dataset-6.
The data in dataset-2 is collected when the participant is sitting.
Dataset-6 contains data on standing, walking, and running.
We take turns selecting one posture for training and testing the other postures for each participant.
For example, we train on sitting posture data and test on sitting, standing, walking, and running posture data.
Similarly, we train on the other three postures.
Fig.~\ref{fig:posture} shows the BAC of SCR-Auth under different postures.
For example, when we use sitting data for training and testing on the rest of the data, the BACs for the four postures are 97.28\%, 95.11\%, 94.43\%, and 91.55\%, respectively.
As observed, the highest BAC is achieved when the posture during both training and testing remains the same. 
The results further reveal that SCR-Auth performs better in sitting, standing, and walking postures than in running. 
In fact, receiving a call while running is relatively uncommon. 
Excluding the 'running' condition, SCR-Auth achieves a BAC of over 94\% in all other postures, underscoring its applicability across diverse postures.

\textbf{Performance over time.}
In this experiment, dataset-2 is used for training, while testing is performed on both dataset-2 and dataset-5. 
Specifically, data in dataset-2 is collected during the first week, and data from the subsequent two weeks is included in dataset-5. 
For weeks 2 and 3, the BACs are 95.77\% and 95.42\%, while the EERs are 4.51\% and 4.78\%, respectively. 
Compared to week 1, the BACs for week 2 and week 3 show slight drops of 1.18\% and 1.53\%. 
This decrease may be attributed to changes in users' postures while holding the device. 
To address this issue, SCR-Auth can be designed to update the authentication model using newly collected data, which is known as the model updating mechanism~\cite{wu2021toward}.

\textbf{Impact of different devices.}
We collected data from three smartphones to evaluate the system performance on different devices.
In this experiment, we use dataset-2 and dataset-7 for evaluation.
Dataset-2 is collected using the Google Pixel 3a, while dataset-7 contains data collected from the Google Pixel 4 and Vivo S12. 
For each user, a one-class SVM classifier is trained on data from a different smartphone.
As shown in Table~\ref{tab:device}, the mean BACs for the Pixel 3a, Pixel 4, and Vivo S12 are 97.32\%, 97.62\%, and 95.20\%, respectively. 
The corresponding average EERs for these devices are 0.85\%, 0.86\%, and 4.03\%.
The results indicate the effectiveness of our system across different devices.

\begin{figure}[!t]
	\centering
	\subfloat[Zero-effort attack]{\includegraphics[width = 0.47\linewidth]{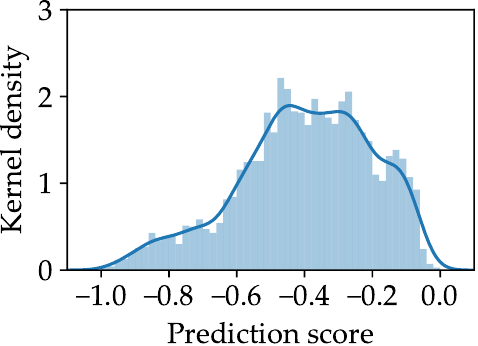}
		\label{fig:attack2}}
	\hspace{0.1em}
	\subfloat[Mimicry attack]{\includegraphics[width = 0.47\linewidth]{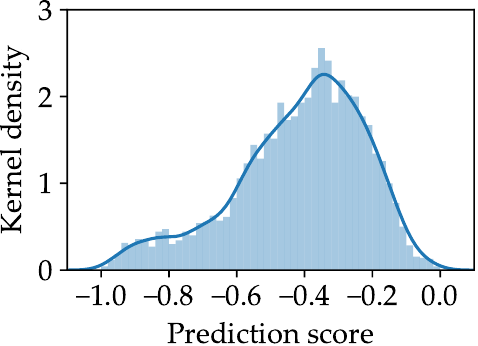}
		\label{fig:attack1}}
	\caption{The kernel density of attack dataset’s prediction score.}
	\label{fig:attack}
\end{figure}

\subsection{Evaluation of Attack Resistance}
To evaluate the system security against two different attacks, we use dataset-8A and dataset-8B to test the authentication model trained using dataset-2.
We use FAR as the rate of wrongly accepted illegal samples in this evaluation.
We also investigate the distribution and kernel density of the attack dataset’s prediction scores, which are evaluated under the Gaussian kernel.

Table~\ref{tab:attack} shows the result of bypassed samples, FAR(\%), and mean prediction scores under two types of attacks.
We test 3500 illegal samples for the zero-effort attack, 33 of which are wrongly accepted.
For the zero-effort attack, the FAR is 0.94\%, and the mean prediction score is -0.39.
We also test on 3500 illegal samples for the mimicry attack, where the number of bypassed samples is 49.
For the mimicry attack, the FAR is 1.40\%, and the mean prediction score is -0.42.
These results demonstrate that SCR-Auth can defend against these two attacks.
The distribution and kernel density of the two attacks' prediction scores are shown in Fig.~\ref{fig:attack}.
Specifically, Fig.~\ref{fig:attack}~\subref{fig:attack2} shows the prediction scores’ distribution under zero-effort attack using dataset-8A.
The kernel density shows a wide range but with a low prediction score.
Fig.~\ref{fig:attack}~\subref{fig:attack1} shows the prediction scores’ distribution under mimicry attack using dataset-8B.
The kernel density shows a wide range and a lower score than the scores under the zero-effort attack.

\subsection{Authentication Latency}
We define the authentication latency of our system as the time from recording the received signal to producing the authentication result.
Therefore, it consists of time for three modules: data preprocessing, feature extraction, and classification.
We developed a prototype system named SCR-Auth on Android to evaluate the authentication latency.
We evaluate one sensing process and compute the average latency from 50 tries.
On Google Pixel 3a, the average authentication latency for the three modules is 82.8ms, 57.6ms, and 69.6ms, respectively.
In total, SCR-Auth requires 0.21s to complete authentication.

\section{Discussion}
\label{sec:discussion}
In this section, we discuss the limitations of our work and experiments, and provide an outlook for potential improvements in future work.

Our study focuses on a realistic call-answering scenario, where call receivers typically use the earpiece mode on smartphones to ensure privacy.
We acknowledge that SCR-Auth does not encompass all usage scenarios.
SCR-Auth may reject a legitimate user who registers with one ear but attempts authentication with the other.
To address this limitation, future work will aim to extend SCR-Auth to support authentication using both ears and accommodate multiple legitimate users.

Due to device limitations, we have not yet evaluated the performance of SCR-Auth against fake ear attacks, which involve using 3D printing technology to create artificial ears for spoofing. 
However, previous studies have highlighted structural discrepancies between 3D-printed ear models and real human anatomy, as well as material differences between fake ear models and human tissue~\cite{Gao2019,xu2024aface}. 
Future research could investigate the impact of such attacks to further enhance the system's security.

While our experiments involved a group of students and three smartphones, larger-scale and multi-device testing is essential to confirm SCR-Auth's applicability in diverse real-world scenarios. 
We recognize that a larger sample size would lead to more robust and generalizable results. 
In future work, we plan to incorporate a broader range of participants and smartphones to further validate the effectiveness and adaptability of SCR-Auth.

\section{Conclusion}
\label{sec:conclusion}
In this paper, we propose SCR-Auth, a secure and implicit call receiver authentication scheme for smartphones that leverages outer ear echoes.
SCR-Auth utilizes the earpiece speaker to emit inaudible sensing signals and the top microphone to record echoes.
In particular, we propose a specially designed two-step denoising method that combines bandpass filtering with magnitude-phase spectrogram subtraction (MPSS) to effectively suppress unwanted interference.
Furthermore, we design a learning-based feature extractor to counteract the position variability, while a one-class classifier is used to verify the legitimacy of the call receiver.
Comprehensive experiments demonstrate that SCR-Auth achieves an average balanced accuracy of 96.95\% and can defend against zero-effort and mimicry attacks.

\bibliographystyle{IEEEtran}
\bibliography{IEEEabrv,earpass}

\end{document}